\documentstyle[11pt,newpasp,twoside,epsf]{article}
\def\edcomment#1{\iffalse\marginpar{\raggedright\sl#1\/}\else\relax\fi}
\reversemarginpar

\begin{document}
\title{Millisecond Pulsars in X-Ray Binaries}
 \author{Deepto Chakrabarty}
\affil{Department of Physics and Center for Space Research, 
Massachusetts~Institute~of~Technology, Cambridge, MA 02139}

\begin{abstract}
Despite considerable evidence verifying that millisecond pulsars are
spun up through sustained accretion in low-mass X-ray binaries
(LMXBs), it has proven surprisingly difficult to actually detect
millisecond X-ray pulsars in LMXBs.  There are only 5
accretion-powered millisecond X-ray pulsars known among more than 80
LMXBs containing neutron stars, but there are another 11
``nuclear-powered'' millisecond pulsars which reveal their spin only
during brief, thermonuclear X-ray bursts.  In addition, 2 of the
accretion-powered pulsars also exhibit X-ray burst oscillations, and
their unusual properties, along with the absence of persistent
pulsations in most LMXBs, suggest that the magnetic fields in many
LMXBs may be hidden by accreted material.  Interestingly, the
nuclear-powered pulsars offer a statistically unbiased probe of the
spin distribution of recycled pulsars and show that this distribution
cuts off sharply above 730~Hz, well below the breakup spin rate for
most neutron star equations of state.  This indicates that some mechanism
acts to halt or balance spin-up due to accretion and that
submillisecond pulsars must be very rare (and are possibly
nonexistent).  It is unclear what provides the necessary angular
momentum sink, although gravitational radiation is an attractive
possibility.
\end{abstract}

\section{Introduction}

Since the discovery of the first millisecond radio pulsar (Backer et
al.  1982), it has been believed that these old, weak-field ($\sim
10^8$ G) pulsars were spun up to millisecond periods by sustained
accretion onto neutron stars in low-mass X-ray binaries (NS/LMXBs;
Alpar et al. 1982).  It is known from accretion torque theory
(e.g., Ghosh \& Lamb 1979) and from observations of strong-field
($>10^{11}$ G) accreting neutron stars (Bildsten et al. 1997) that
steady disk accretion onto a magnetized neutron star will lead to an
equilibrium spin period
\begin{equation}
P_{\rm eq} \sim 1\mbox{\rm\ s\,}
           \left(\frac{B}{10^{12}\mbox{\rm\,G}}\right)^{6/7}
           \left(\frac{\dot M}{10^{-9} M_\odot\mbox{\rm\,yr$^{-1}$}}
            \right)^{-3/7},
\end{equation}
where $B$ is the pulsar's surface dipole magnetic field strength and 
$\dot M$ is the mass accretion rate.  Also, most neutron stars in
LMXBs are old and have weak magnetic fields, based both on the
occurrence of thermonuclear X-ray bursts (which require $B<10^{10}$ G;
see Joss \& Li 1980) and on their X-ray spectral (Psaltis \& Lamb
1998) and timing properties (see, e.g., van der Klis 2000).  Thus, as
long as we accept that sustained accretion onto neutron stars can somehow
attenuate their strong birth fields down to $\sim 10^8$~G strengths
(see, e.g., Bhattacharya \& Srinivasan 1995), it is natural to expect
neutron stars in LMXBs to be spinning at millisecond periods.  (Note
that although the general trend of Equation~(1) is expected to extend
down to $10^8$~G, the precise power-law dependences are likely to be
modified; see Psaltis \& Chakrabarty 1999).  

A robust prediction of this model is that the neutron stars in LMXBs
should be X-ray pulsars, since a $10^8$~G field should be strong
enough to truncate the Keplerian accretion disk and channel its flow
onto the neutron star's magnetic poles.  However, the detection of
accretion-powered millisecond X-ray pulsars proved elusive for nearly
two decades, with a series of X-ray missions failing to detect
millisecond pulsations from NS/LMXBs down to stringent upper limits on
the pulsed fraction (e.g., Vaughan et al. 1994).  The
launch of the {\em Rossi X-Ray Timing Explorer} ({\em RXTE}; Bradt,
Rothschild, \& Swank 1993; Jahoda et al. 1996) in December
1995 finally provided an instrument with sufficient flexibility and
sensitivity to detect SAX J1808.4$-$3658, the first accretion-powered
millisecond X-ray pulsar (Wijnands \& van der Klis 1998; Chakrabarty
\& Morgan 1998), as well as several additional examples (see
Table~1).  A key element proved to be the highly flexible pointing and
scheduling ability of {\em RXTE}.  The Galactic X-ray sky is highly
variable, with some sources lying dormant for years and only
intermittently becoming active in X-ray emission.   The combination of
having both the {\em RXTE} All Sky Monitor to determine when an X-ray transient
becomes active and the ability to rapidly repoint the main {\em RXTE}
instruments at a newly active source were crucial in enabling the
discovery of millisecond X-ray pulsars.  

Along the way, {\em RXTE} detected other classes of rapid X-ray
variability which also point to millisecond spins for these neutron
stars but which also raised a number of new questions about the
underlying physics.  

\section{Millisecond Variability in Accreting Neutron Stars}

{\em RXTE} has identified three distinct classes of millisecond
variability in accreting neutron stars:
\begin{itemize}
  \item {\bf Kilohertz quasi-periodic oscillations (kHz QPOs)} (van
  der Klis et al. 1996; see van der Klis 2000 for a review): These are
  pairs of relatively high-$Q$ peaks in the X-ray intensity power
  spectrum whose frequencies drift by hundreds of hertz (200--1200 Hz)
  as the source intensity varies, but whose separation frequency
  (typically a few hundred Hz) remains roughly constant (see example
  in left panel of Figure~1).  This approximate separation frequency
  has a unique, reproducible value for each of the over 20 NS/LMXBs in
  which this phenomenon is observed.  These oscillations are believed
  to arise in the inner accretion disk flow, where the dynamical time
  scale is of order milliseconds.
\begin{figure}[t]
%\plottwo{khzscox1.ps}{burst1702.ps}
\plottwo{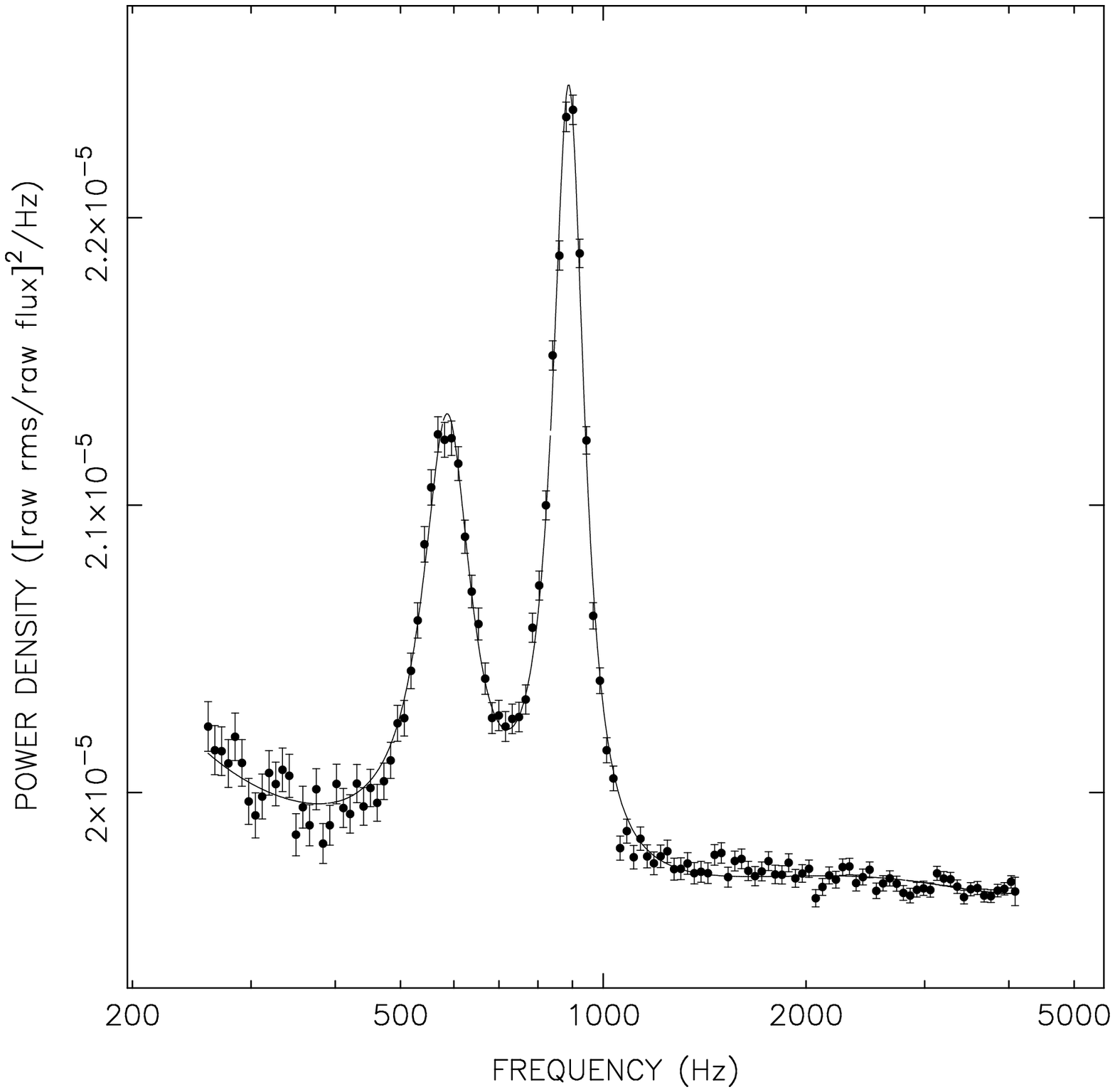}{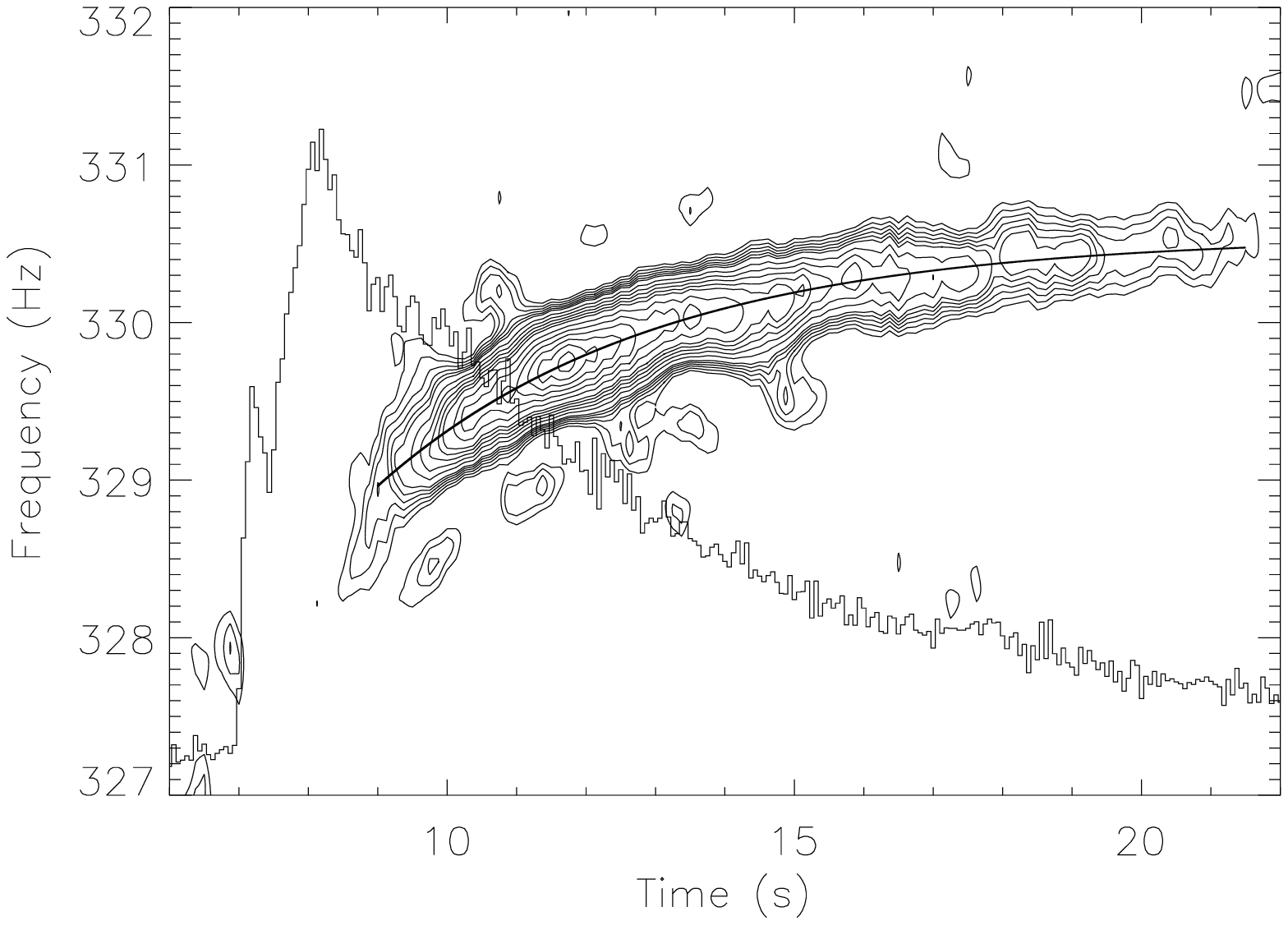}
\caption{{\em Left:} Power spectrum of X-ray intensity showing kHz QPO
  pair in Sco 
  X-1. Adapted from van der Klis et al. (1997). {\em Right:} X-ray
  burst oscillations in 4U 1702$-$43. The solid histogram shows the
  X-ray intensity history of the X-ray burst.  The contours show the
  Fourier power level as a function of frequency and time, indicating
  a drifting oscillation starting at 328~Hz ($t=7$~s) and ending at 
  330.5~Hz. Adapted from Strohmayer \&  Markwardt (1999).}
\end{figure}

  \item {\bf X-ray burst oscillations} (Strohmayer et al. 1996; see
  Strohmayer \& Bildsten 2004 for a review): These are nearly coherent
  millisecond oscillations observed only during thermonuclear X-ray
  bursts (which typically last $\sim$10~s).  The observed frequencies
  drift by a few Hz over $\sim$5~s during the bursts, asymptotically
  reaching a maximum frequency that is unique and reproducible for
  each of the 13 sources in which this phenomenon is observed (see
  right panel of Figure~1).  These maximum frequencies lie in the
  270--619~Hz range. The oscillation (at least at the burst onset) is
  understood as a temperature anisotropy caused by the nuclear
  burning, with the frequency drift interpreted as angular momentum
  conservation in a cooling, decoupled burning layer on the stellar
  surface (Strohmayer et al. 1997; Cumming \& Bildsten 2000).  Since
  their discovery, it was suspected that the millisecond oscillations
  were somehow tracing the stellar spin, but until recently there was
  some question about whether the oscillations might be a harmonic of
  the spin.  A list of these sources is given in Table~1.

  \item {\bf Persistent accretion-powered pulsations} (Wijnands \& van
  der Klis 1998; see Wijnands 2004 for a review): These are the
  objects originally expected by the recycling hypothesis, NS/LMXBs whose
  persistent (non-burst) emission contains coherent millisecond
  pulsations (see example in Figure~2).  Oddly, all five of the these
  pulsars (see Table~1) are in soft X-ray transients: accreting
  systems that lie dormant for years, with intermittent outbursts of
  accretion lasting about a month (these outbursts are understood to
  arise from an accretion disk instability; see Frank, King, \& Raine
  2002).  Moreover, all five also have very short orbital
  periods\footnote{As an aside, I also note that there are five LMXBs
  whose orbital periods are within 90 seconds of 42 minutes: three of
  the millisecond X-ray pulsars, the strong-field pulsar 4U 1626$-$67, and
  the quiescent LMXB NGC 6652B.  This excess is statistically
  significant, and there is no selection effect for finding this
  period.  What is so special about this particular orbital period?}
  and very low mass accretion rates.  Also, the pulsed amplitude of
  these systems is about 6 percent, while the upper limit on the
  pulsed amplitude for most NS/LMXBs is 1 per cent or less, raising
  the question of what is different about this group of 5 pulsars: why
  is it so difficult to find persistent millisecond pulsations in most
  NS/LMXBs?  It should be noted that none of the millisecond X-ray
  pulsars has been detected as a radio pulsar.
\end{itemize}
\begin{figure}[t]
\plotfiddle{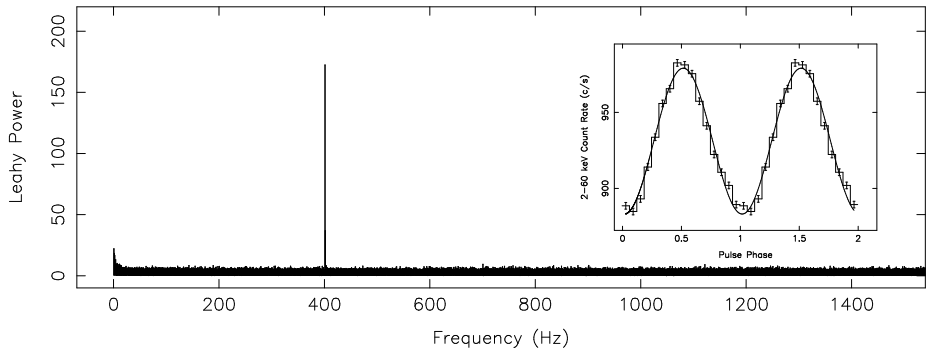}{2in}{270}{50}{50}{-200}{220}
\caption{Fourier power spectrum showing coherent 401~Hz X-ray
  pulsations in the persistent pulsar SAX J1808.4$-$3658. The inset
  shows the highly sinusoidal pulse profile. Adapted from Wijnands \&
  van der Klis (1998).} 
\end{figure}

While both kHz QPOs and X-ray burst oscillations are often observed
in the same source, until 2002 neither phenomena had ever been 
observed in accretion-powered millisecond pulsar.  Since it was only
in this latter class that the neutron star spin was definitively
known, the relationship between these three classes of variability was
not clear.  Still, some patterns were clear.  In particular, the
separation frequency $\Delta\nu_{\rm kHz}$ organized the burst
oscillation sources into two groups: the slow oscillators (with
$\nu_{\rm burst}<400$~Hz) all have $\nu_{\rm burst}\approx
\Delta\nu_{\rm kHz}$, while the fast oscillators (with $\nu_{\rm
burst}>500$~Hz) all have $\nu_{\rm burst}\approx 2\Delta\nu_{\rm
kHz}$.  The photospheric radius expansion properties of the X-ray
bursts also divide along these lines, with fast oscillations occurring
preferentially in radius expansion bursts (Muno, Galloway, \&
Chakrabarty 2004).  Since both $\Delta\nu_{\rm kHz}$ and $\nu_{\rm
burst}$ are reproducible characteristics of a given source and the
most likely mechanism for such a stable frequency is the stellar spin,
there was considerable debate as to whether it is $\Delta\nu_{\rm
kHz}$ or $\nu_{\rm burst}$ that is the fundamental spin frequency.

This question was finally settled by the detection in kHz QPOs in two
accretion-powered X-ray pulsars in 2002 and 2003.  In the rapid rotator SAX 
J1808.4$-$3658, $\Delta\nu_{\rm kHz}$ is half the 401~Hz spin
frequency (Wijnands et al. 2003); while in the slow rotator XTE
J1807$-$294, $\Delta\nu_{\rm kHz}$ roughly equals the 190~Hz spin
frequency (Markwardt et al. 2004), verifying the odd phenomenology
described above.  (The possibility that the spin frequency in SAX
J1808.4$-$3658 is actually 200.5~Hz was excluded by the very stringent
non-detection of pulsations at this frequency; see Morgan et
al. 2004.) In addition, X-ray burst oscillations were also detected in
the 401~Hz pulsar SAX J1808.4$-$3658 (Chakrabarty et al. 2003) and the
314~Hz pulsar XTE J1814$-$338 (Strohmayer et al. 2003); in both cases,
$\nu_{\rm burst}$ was equal to the spin frequency.    

These new observations lead to three conclusions:
\begin{itemize}
  \item X-ray burst oscillations directly trace the neutron star spin
  (and are not higher harmonics of a fundamental), and may thus be
  thought of as nuclear-powered pulsations.

  \item The kHz QPO separation frequency $\Delta\nu_{\rm kHz}$ is
  sometimes roughly the spin frequency (in slow rotators) and
  sometimes roughly half the spin frequency (in fast rotators).  The
  origin of these QPOs is uncertain, but the underlying mechanism
  clearly has some coupling to the stellar spin.

  \item Most neutron stars in LMXBs are indeed spinning at millisecond
  periods, but for some reason only a small fraction of them are
  visible as persistent, accretion-powered millisecond pulsars.
\end{itemize}

This last point is puzzling, since one would expect a $10^8$~G
accreting neutron star to be a pulsar.  There have been several
possible explanations discussed for why most NS/LMXBs are not
pulsars.  One possibility is that the most of the NSs have magnetic
fields that are too weak to channel the accretion flow, although there
is evidence for radio pulsars with fields much weaker than $10^8$~G.
Another possibility is that the neutron stars in non-pulsing LMXBs are
surrounded by a scattering medium that attenuates pulsations (Brainerd
\& Lamb 1987; Kylafis \& Phinney 1989).  Indeed, Titarchuk, Cui, \&
Wood (2003) have recently argued that the data support this argument,
although there is not yet a consensus on this point (Heindl \& Smith
1998; Psaltis \& Chakrabarty 1999).  A third possibility is that
gravitational self-lensing might attenuate the pulsations (Wood,
Ftaclas, \& Kearney 1988; Meszaros, Riffert, \& Berthiaume 1988).  In
the next section, I will discuss recent evidence that suggests that
magnetic field strength may indeed be the relevant factor.

\section{SAX J1808.4$-$3658: Evidence for a Range of LMXB Magnetic
  Field Strengths}

It is instructive to compare the behavior of the X-ray burst
oscillations observed in the pulsar SAX J1808.4$-$3658 with those
observed in the non-pulsing LMXBs.  In SAX J1808.4$-$3658,
strong millisecond oscillations around the 401~Hz spin frequency were
observed during 4 X-ray bursts in 2002, with very similar
characteristics in each burst (Chakrabarty et al. 2003).  An example
is shown in Figure~3.  First, a rapidly drifting oscillation
(increasing from 397 to 403 Hz) was detected during the burst
rise. Second, no oscillations were detected during the radius expansion
phase of the burst (typical of other burst oscillation sources as
well).  Finally, a strong oscillation reappeared during the cooling
phase of the burst, at a constant frequency nearly equal to the spin
frequency (which was known precisely from the pre-burst persistent
pulsations), but exceeding it by one part in 70000.

\begin{figure}[t]
%\plotfiddle{1808burst.ps}{3in}{0}{70}{70}{-220}{-210}
\plotfiddle{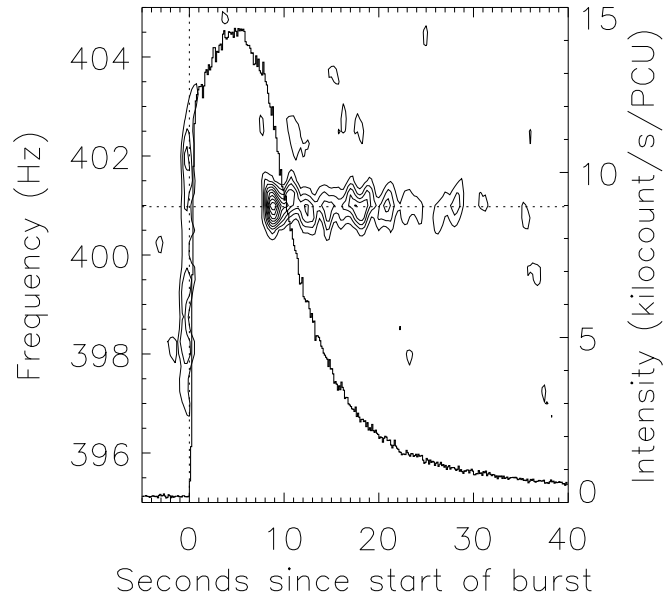}{3in}{0}{70}{70}{-220}{-210}
\caption{X-ray burst oscillation in the pulsar SAX J1808.4$-$3658. The
histogram shows the X-ray intensity during the thermonuclear X-ray
burst.  The contours show the Fourier power levels as a function of
frequency and time. The horizontal dotted line indicates the (known)
pulsar spin frequency.  A rapidly drifting oscillation is detected
during the burst rise and overshoots the spin rate.  A stationary
oscillation near the spin rate is detected in the burst tail. Adapted
from Chakrabarty et al. (2003).}
\end{figure}

This observed frequency drift demonstrates that this is a similar
phenomena as the burst oscillations observed in other (non-pulsing)
neutron stars, although the oscillations in SAX J1808.4$-$3658 have
some very unusual traits: the drift time scale is an order of
magnitude faster than in the other neutron stars (compare right panel
of Figure~1), and the maximum oscillation frequency is reached during
the burst rise, inconsistent with angular momentum conservation in a
cooling, contracting burning shell.  In fact, as evident in Figure~3,
the oscillation {\em overshoots} the spin frequency during the burst
rise.  The rapid, overshooting drift probably indicates that SAX
J1808.4$-$3658 has a stronger magnetic field than the other burst
oscillation sources (Chakrabarty et al. 2003), since a sufficiently
strong field will suppress rotational shearing in the burning layer
and may act as a restoring force.  Strohmayer et al. (2003) also
interpreted the frequency evolution of the burst oscillations in the
pulsar XTE J1814$-$338 as evidence for a stronger than normal magnetic
field in that LMXB.

This magnetic field argument is particularly appealing given that only
SAX J1808.4$-$3658 and XTE J1814$-$338, among a total of 13 burst
oscillation sources (and two of only five systems out of a total of
over 80 NS/LMXBs), show persistent pulsations in their non-burst
emission.  The absence of persistent pulsations in most of these
systems suggests that they lack a sufficiently strong magnetic field
for the accretion flow to be magnetically channeled.  Indeed, it has
been proposed that the absence of persistent pulsations in most
NS/LMXBs is due to diamagnetic screening of the neutron star magnetic
field by freshly accreted material, which would occur above a critical
value of $\dot M$ at a few percent of the Eddington rate (Cumming,
Zweibel, \& Bildsten 2001).  In this context, it is interesting to
note that all five of the persistent millisecond X-ray pulsars like at
the low end of the $\dot M$-distribution for LMXBs.  Thus, while all
the NS/LMXBs may have underlying surface field strengths of $\sim
10^8$~G (see, e.g., Psaltis \& Chakrabarty 1999), it may be that the
non-pulsing LMXBs have {\em effective} field strengths that are much lower due
to screening by accreted material.  Presumably, the underlying field
would emerge when the accretion eventually halts; thus,
this idea is still consistent with the absence of millisecond radio
pulsars with fields much weaker than $10^8$~G.

\section{The Underlying Spin Distribution of Recycled Pulsars}

Having observationally verified that millisecond pulsars are spun up
or ``recycled'' in LMXBs, it is interesting to ask what the underlying
spin distribution of recycled pulsars is, and whether there is any
limit to the recycling process.  One might expect the distribution of
spin frequencies to simply reflect the range of equilibrium spins
corresponding to the magnetic field strength distribution in LMXBs.
However, if the effective field strength of most NS/LMXBs is
considerably below $10^8$~G as discussed above, then the resulting
boundary layer accretion onto the neutron star might be capable of
spinning up the pulsar to submillisecond periods.  Certainly, a strict
upper limit on the neutron star spin rate is given by the centrifugal
breakup limit, up to 3~kHz depending upon the neutron star equation of
state (Cook, Shapiro, \& Teukolsky 1994; Haensel, Lasota, \& Zdunik
1999). 

\begin{figure}[t]
%\plotone{fhist.ps}
\plotone{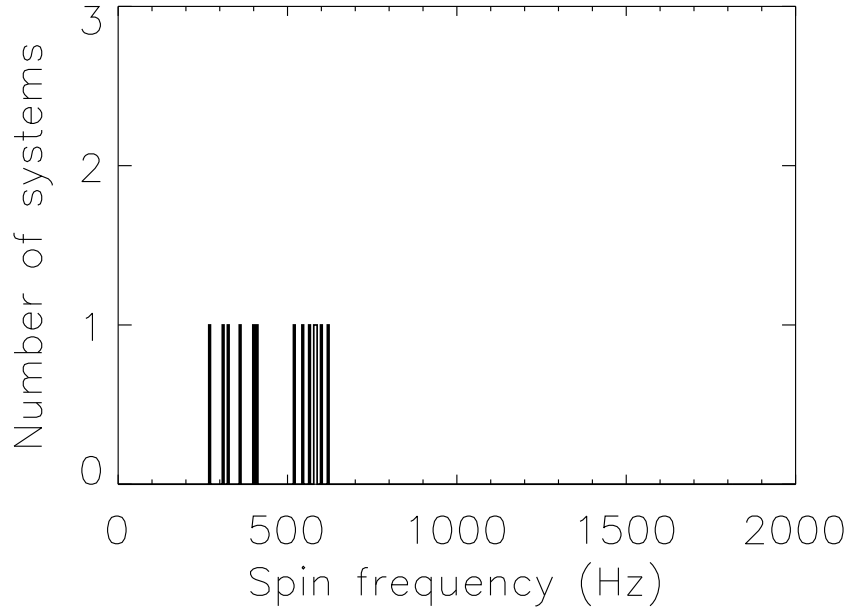}
\caption{The spin frequency distribution of nuclear-powered
  millisecond X-ray pulsars.  There is a sharp drop off in the
  population at spins above 730~Hz.  {\em RXTE} has no significant
  selection effects against detecting oscillations as fast as 2000~Hz,
  making the absence of fast rotators extremely statistically significant.
  Based on results from Chakrabarty et al. (2003).}
\end{figure}

Although the substantial known population of millisecond radio pulsars
in principle provides an ideal probe of the spin distribution of
recycled pulsars, severe observational selection effects have
historically made it difficult to make a statistically accurate
estimate.  However, the burst oscillation sources (nuclear-powered
pulsars) are an ideal probe: they are bright and easily detected
throughout the Galaxy, their signals are short-lived enough to avoid modulation
losses due to orbital Doppler smearing, and {\em RXTE} has no
significant selection effects against detecting oscillations as fast
as 2 kHz.  

The spin frequencies (see Table~1) of the 13 known nuclear-powered
millisecond pulsars are plotted in Figure~4.  (The three non-bursting,
accretion-powered millisecond pulsars are omitted to keep the sample
unbiased.)  The spins are consistent with a uniform distribution
within the observed 270--619~Hz range.  The absence of any pulsars at
lower frequencies is not surprising, since such low equilibrium spin
rates would require somewhat higher magnetic field strengths, which
would then suppress the thermonuclear X-ray bursts necessary for burst
oscillations.  However, the absence of spins above 619~Hz is extremely
significant, given that there is no significant loss of {\em RXTE}
sensitivity out to 2~kHz.  Under the simple assumption of a uniform
distribution out to some maximum value, the observed distribution
yields an maximum spin frequency of 730~Hz (95\% confidence;
Chakrabarty et al. 2003).  This limit is consistent with the fastest
known millisecond radio pulsar, PSR B1937+21, which has $P_{\rm spin}
= 641$~Hz.  This limit is also well below the breakup frequency for
nearly all equations of state for rapidly rotating neutron stars.
Recent radio pulsar surveys, in which selection effects are accounted
for, are independently finding similar evidence for a maximum spin
frequency around 700~Hz, as reported at this meeting (McLaughlin et
al. 2004; Camilo 2004). 

\begin{table}[t]
\caption{Millisecond Pulsars in X-Ray Binaries}
\begin{tabular}{lccl}
\tableline
Object & Spin Frequency (Hz) & Orbital Period \\
\tableline
\multicolumn{3}{l}{ACCRETION-POWERED PULSARS} \\
XTE J0929$-$314 & 185 & 43.6 min \\
XTE J1807$-$294 & 191 & 41 min  \\
XTE J1814$-$338 & 314 & 4.27 hr  \\
SAX J1808.4$-$3658 & 401 & 2.01 hr  \\
XTE J1751$-$305 & 435 & 42.4 min  \\
 & & \\
\multicolumn{3}{l}{NUCLEAR-POWERED PULSARS (Burst Oscillations)} \\
4U 1916$-$05 & 270  & 50 min\\
XTE J1814$-$338 & 314 & 4.27 hr\\
4U 1702$-$429 & 330 & ?\\
4U 1728$-$34 & 363 & ?\\
SAX J1808.4$-$3658 & 401 & 2.01 hr\\
SAX J1748.9$-$2021 & 410 & ?\\
KS 1731$-$260 & 524 & ?\\
Aql X-1 & 549 & 19.0 hr\\
X1658$-$298 & 567 & 7.11 hr\\
4U 1636$-$53 & 581 & 3.8 hr\\
X1743$-$29 & 589 & ?\\
SAX J1750.8$-$2900 & 601 & ?\\
4U 1608$-$52 & 619 & ?\\
\tableline
\end{tabular}
\end{table}

It is thus clearly demonstrated observationally that the population of
pulsars with spins above 730~Hz must drop off dramatically (although
not necessarily to zero: the existence of submillisecond pulsars is
{\em not} excluded, but such objects are evidently at best very rare).
It remains unclear what causes this drop off.  Magnetic spin
equilibrium can only account for the observed distribution if the entire
sample has surface magnetic field strength $\sim 10^8$~G.  However, as
noted above, fields this strong should be dynamically important for the
accretion flow and lead to persistent millisecond pulsations, making
it difficult to understand the lack of pulsations in most NS/LMXBs and
instead suggesting a wider range of magnetic field strengths.  

Alternatively, several authors have shown that gravitational radiation
can carry away substantial angular momentum from accreting neutron
stars, driven by excitation of an $r$-mode instability in the neutron
star core (Wagoner 1984; Andersson, Kokkotas, \& Stergioulas 1999), or
by a rotating, accretion-induced crustal quadrupole moment (Bildsten
1998), or by large (internal) toroidal magnetic fields (Cutler 2002).
These losses can balance accretion torques for the relevant ranges of
spin and $\dot M$, and the predicted gravitational radiation strengths
are near the detection threshold for, e.g., the planned Advanced LIGO
detectors (Bildsten 2003).  Such a detection would rely on operating a
narrow-band search in a known frequency range, but this approach has
the advantage of searching for a persistent signal that can be
integrated, unlike the transient signals like those produced by binary
mergers. 

It will be of considerable interest to measure the shape of the spin
distribution more accurately, in order to determine whether we are
observing a gradual fall off, an abrupt cutoff, a pile up, etc. --- at
present, the data cannot distinguish between these possibilities, and
thus cannot meaningfully discriminate between theoretical models.  In
the long run, it is likely to be the radio systems that provide the
best measurement of the distribution, based simply on ease of detectability
and the available instrumentation.

\acknowledgements{It is a pleasure to thank Lars Bildsten, Duncan
  Galloway, Scott Hughes, Fred Lamb, Craig Markwardt, Nergis
  Mavalvala, Ed Morgan, Mike Muno, Dimitrios Psaltis, Tod Strohmayer,
  Michiel van der Klis, and Rudy Wijnands for useful discussions and
  collaborations.  I also thank Fred Rasio, Ingrid Stairs, and Steve
  Thorsett for inviting me to give this review at Aspen.}

\end{document}